\documentclass[12pt]{article}

\newcommand{\be}{\begin{equation}}\newcommand{\ee}{\end{equation}}
\newcommand{\bea}{\begin{eqnarray}}\newcommand{\eea}{\end{eqnarray}}
\newcommand{\nn}{\nonumber}\newcommand{\p}[1]{(\ref{#1})}
\newcommand{\lb}[1]{\label{#1}}

\newcommand\q{\quad}

\newcommand{\vp}{\varphi}
\newcommand{\bvp}{{\bar\varphi}}
\newcommand{\bnu}{{\bar\nu}}

\newcommand\cA{{\cal A}}

\newcommand\cL{{\cal L}}

\newcommand\cS{{\cal S}}

\newcommand{\da}{{\dot{\alpha}}}
\newcommand{\db}{{\dot{\beta}}}

\newcommand\ab{{\alpha\beta}}

\newcommand\adb{{\alpha\db}}
\newcommand\ada{{\alpha\da}}

\newcommand\padb{\partial_\adb}

\begin{document}

\begin{center}
{\bf Invariance of interaction terms in new representation \\
of self-dual electrodynamics}\\
\vspace{0.5cm}

{\bf B.M. Zupnik}\\
\end{center}

{Bogoliubov Laboratory of Theoretical Physics, Joint Institute for
Nuclear Research, Dubna, Moscow Region, 141980, Russia; e-mail: 
zupnik@thsun1.jinr.ru}

\begin{abstract}
 A new representation of Lagrangians of $4D$ nonlinear
electrodynamics is considered.
In this new formulation, in parallel with the standard Maxwell field
strength $F_{\alpha\beta}, \bar{F}_{\dot\alpha\dot\beta}$, an auxiliary
bispinor field $V_{\alpha\beta}, \bar{V}_{\dot\alpha\dot\beta}$ is
introduced. The gauge field strength appears only in bilinear terms
of the full Lagrangian, while the interaction Lagrangian $E$ depends on
the auxiliary fields, $E = E(V^2, \bar V^2)$.  Two types of self-duality 
inherent in the nonlinear
electrodynamics models admit a simple characterization in terms of the
function $E$. The continuous $SO(2)$ duality symmetry between nonlinear
equations of motion and Bianchi identities amounts to requiring $E$ to be
a function of the $SO(2)$ invariant quartic combination $V^2\bar V^2$.
The discrete self-duality (or self-duality under
Legendre transformation) amounts to a weaker condition $E(V^2, \bar{V}^2)
= E(-V^2, -\bar{V}^2)$.  This approach can be generalized to a
system of $n$ Abelian gauge fields exhibiting $U(n)$ duality. The
corresponding interaction Lagrangian should be $U(n)$ invariant function
of $n$ bispinor auxiliary fields.
\end{abstract}

\section{Introduction}

This talk is based on our paper \cite{IZ2} (see, also Sec.3 of 
ref.\cite{IZ}). It is well known that the on-shell $SO(2)$ ($U(1)$) duality 
invariance of Maxwell equations can be generalized to the whole class of the 
nonlinear electrodynamics models, including the famous Born-Infeld theory. 
The condition of $SO(2)$ duality can be formulated as a nonlinear 
differential equation for the Lagrangian of these theories \cite{GZ,GR,KT}. 
This self-duality equation is equivalent to the known Courant-Gilbert 
equation, and its general solution can be reduced to an algebraic equation
for the auxiliary real scalar variable and depends on an arbitrary real
function of this variable \cite{GZ,HKS}.

Note that only few examples of the exact solutions for self-dual Lagrangians 
$L(F_{mn})$ are known. The general self-dual solution can be analyzed in the
representation with an additional auxiliary scalar field \cite{HKS}, however,
the symmetry properties of this representation have not been studied.

We shall consider the new representation of self-dual electromagnetic
Lagrangians using the auxiliary bispinor (tensor) fields which transform
linearly with respect to the $U(1)$ duality group. In this representation, 
{\it the self-duality condition is equivalent to the $U(1)$-invariance of the 
nonlinear self-interaction of auxiliary fields}, while the bilinear free
Lagrangian is not invariant. The Lagrangian involving only the Maxwell 
field strengths emerges as a result of eliminating the bispinor auxiliary 
fields by their algebraic equations of motion. More general nonlinear 
electrodynamics Lagrangians respecting the so-called discrete self-duality
(or duality under Legendre transformation) also admit a simple 
characterization in terms of the self-interaction of the bispinor auxiliary 
fields. In this case it should be invariant with respect to a discrete 
self-duality transformation. 

The linearly transforming bispinor auxiliary fields are very useful in the 
analysis of $U(n)$ duality-invariant systems of $n$ Abelian gauge fields.
The corresponding Lagrangian  is fully specified by the interaction term 
which is an $U(n)$ invariant function of $n$ auxiliary fields. The discrete 
self-dualities also amount to a simple restriction on the interaction 
function.

\section{ Self-dualities in nonlinear electrodynamics}

Let us introduce the spinor notation for the Maxwell field strengths
$F_\ab, \bar F_{\da\db}$ and the following the Lorentz-invariant complex 
variables:
\bea
&& \varphi = F^\ab F_\ab~,
\quad \bar\varphi  =
\bar{F}^{\da\db}\bar{F}_{\da\db}~.
\lb{compl}
\eea
In this representation, two independent invariants which one can construct
out of the Maxwell field strength in the standard vector notation take the
following form:
\bea
&&F^{mn}F_{mn}=2( \vp+\bvp)~,\nn\\
&&F_{mn}
\widetilde{F}^{mn} =-2i(\vp-\bvp)~. \lb{vecsp}
\eea
Here
$$
F_{mn}=\partial_mA_n-\partial_nA_m~,\q\widetilde{F}^{mn}
={1\over2}\,\varepsilon^{mnpq}F_{pq}.
$$

It will be convenient to deal with dimensionless $F_{\alpha\beta},
\bar{F}_{\dot\alpha\dot\beta}$ and $\vp, \bvp$, introducing a coupling
constant $f$, $[f] = 2$. Then the generic nonlinear Lagrangian 
is proportional to $f^{-2}$ and contains the dimensionless function
\be
L(\vp,\bvp) =
-{1\over2}(\vp+\bvp)+L_{int}(\vp,\bvp)
\lb{nonl}
\ee
where $L_{int}(\vp, \bvp)$ collects all possible self-interaction terms of
higher-order in $\vp, \bvp$.

We shall use the following notation for the derivatives of the Lagrangian
$L(\vp,\bvp)$ 
\bea
&&P_\ab(F)\equiv i\partial L/\partial F^\ab=2iF_\ab
 L_\vp~,\lb{defP}\\
&&L_\vp= \partial L/\partial \vp~,\q L_\bvp= \partial L/ \partial
\bvp~.\;\nn 
\eea

The nonlinear equations of motion have the following form in this
representation:
\bea
&& {\cal E}_\ada(F)\equiv\partial_\alpha^\db \bar P_{\da\db}(F)
-\partial^\beta_\da P_\ab(F)= 0~, \lb{BIeq}
\eea
where $\padb=(\sigma^m)_\adb\partial_m$.
These equations, together with the Bianchi identities
\bea
&&{\cal B}_\ada(F)\equiv
\partial_\alpha^\db \bar F_{\da\db}-\partial^\beta_\da F_\ab
= 0~, \lb{Bian}
\eea
constitute a set of first-order equations in which one can treat $F_\ab$
and $\bar{F}_{\da\db}$ as {\it unconstrained} conjugated variables.

This set is said to be duality-invariant if the Lagrangian $L(\vp, \bvp)$
satisfies certain nonlinear condition \cite{GZ,GR,KT}. The precise form
of this self-duality condition in the spinor notation is
\bea
&S[F,P(F)]\equiv F^\ab F_\ab+P^\ab P_\ab-\mbox{c.c.}
\nn\\
&=\vp - \bvp - 4\,[\vp(L_\vp)^2 - \bvp(L_\bvp)^2]=0.
\lb{sdI}
\eea
Using this condition, one can  define the nonlinear transformations
\bea
&& \delta_\omega F_\ab=\omega\,P_\ab(F)\equiv 2i\,\omega\,F_\ab L_\vp~,\q
\delta_\omega P_\ab(F)=-\omega F_\ab
\lb{Ftrans}
\eea
where $\omega$ is a real parameter. 
The set of equations \p{BIeq}, \p{Bian} is clearly invariant under
these transformations.

Although the Lagrangian $L(\vp,\bvp)$ satisfying \p{sdI} is not invariant
with respect to transformation \p{Ftrans}, one can still construct the 
$SO(2)$ invariant function
\be
\delta_\omega[ L+{i\over2}(P_\ab F^\ab-\bar{P}_{\da\db}
\bar{F}^{\da\db})]=0~.\lb{invfun}
\ee
The self-duality condition is equivalent to the following relations
for the $SO(2)$ variation of the Lagrangian:
\bea
&&\delta_\omega L\equiv \delta F_\ab\frac{\partial L}{\partial F_\ab}+
\delta\bar F_{\da\db}\frac{\partial L}{\partial \bar F_{\da\db}}=
-i\omega(P^2-\bar P^2)\nn\\
&&={i\over2}\delta_\omega(P_\ab F^\ab-\bar{P}_{\da\db}
\bar{F}^{\da\db})=-i\omega(F^2-\bar F^2)~.\lb{deltaL}
\eea

We shall consider the discrete self-duality condition
as an invariance of the Lagrangian with respect to the Legendre transforms
\bea
&&L(\vp,\bvp) \Rightarrow L(P^2,\bar P^2)
=L(\vp,\bvp)+i(F^\ab P_\ab-\bar{F}^{\da\db}\bar{P}_{\da\db})\lb{discr}\\
&&F_\ab~\Rightarrow~P_\ab=i\partial L/\partial F^\ab~,\q \bar F_{\da\db}
\Rightarrow \bar P_{\da\db}~.\nn
\eea

It is easy to show  that the $SO(2)$ duality condition \p{sdI}  guarantees
the self-duality under Legendre transformation (see, e.g., \cite{KT}).

\section{Linearly transforming auxiliary fields in a new\\
 representation of nonlinear electrodynamics}

The recently constructed $N=3$ supersymmetric extension of the Born-Infeld
theory \cite{IZ} suggests a new representation for the actions of
nonlinear electrodynamics discussed in the previous Section.

Let us consider the auxiliary fields $V_{\alpha\beta}, \bar V_{\da\db}$ and 
the following generalization of the free Lagrangian:
\be
\cL_2(V,F)=\nu+ \bnu- 2\,(V^\ab F_\ab+\bar{V}^{\da\db}\bar{F}_{\da\db})
+{1\over2}(\vp+\bvp)~, \lb{auxfree}
\ee
where $\nu\equiv V^\ab V_\ab$ and $\bnu\equiv\bar{V}^{\da\db}
\bar{V}_{\da\db}$.

Eliminating $V^\ab$ by its algebraic equation of motion,
\bea
&& V^\ab = F^\ab~, \q \bar V^{\da\db} = \bar F^{\da\db}~,\lb{free1}
\eea
we arrive at the free Maxwell Lagrangian.

Our aim will be to find a nonlinear extension of the free Maxwell Lagrangian 
using $\cL_2(V, F)$ , such that this extension becomes the generic
nonlinear Lagrangian $L(F^2,\bar{F}^2)$, eq. \p{nonl}, after eliminating
the auxiliary fields $V_{\alpha\beta}, \bar{V}_{\dot\alpha\dot\beta}$ by
their algebraic ({\it nonlinear}) equations of motion.

By Lorentz covariance, such a nonlinear Lagrangian has the following
general form:
\bea
&& \cL[V,F(A)] = \cL_2[V,F(A)] + E(\nu,\bnu)~, \lb{legact}
\eea
where $E$ is a real function encoding self-interaction. Varying the action
with respect to $V_\ab$, we derive the algebraic relation between $V$ and
$F(A)$ in this formalism
\bea
&& F_\ab(A) = V_\ab(1+ E_\nu) \quad \mbox{and c.c.}~, \lb{FV}
\eea
where $E_\nu\equiv\partial E(\nu,\bnu)/\partial\nu$. This relation is a
generalization of the free equation \p{free1} and it can be used to
eliminate the auxiliary variable $V^\ab$ in terms of
$F^\ab$ and $\bar F^{\da\db}$, $V_\ab~\Rightarrow~V_\ab[F(A)]$
(see eq. \p{VF} below). 
The second equation of motion in this representation, obtained by
varying \p{legact} with respect to $A_\ada$, has the form
\bea
&& \partial^\beta_\da[F_\ab(A)-2V_\ab]+\mbox{c.c.}=0~.\lb{FV3}
\eea
After substituting $V_\ab = V_\ab[F(A)]$ from \p{FV}, eq. \p{FV3} becomes
the dynamical equation for $F_\ab(A), \bar F_{\da\db}(A)$ corresponding
to the generic Lagrangian  \p{nonl}. Comparing \p{FV3} with \p{BIeq}
yields the relation
\bea
&& P_\ab(F)\equiv -2iF_\ab L_\vp = i\left[ F_\ab - 2V_\ab(F) \right]~, 
\lb{imprel}\\
&& V_\ab(F)=F_\ab G(\vp,\bvp)\lb{VF}~,\q G={1\over2}-L_\vp~.\lb{exprG}
\eea

One should also add the relations:
\bea
&& G^{-1}=1+E_\nu~,\q\bar{G}^{-1}=1+E_\bnu~, \lb{GE}
\eea
which follow from  \p{VF}.

A useful corollary of these formulas  is the
relation
\bea
&& \nu E_\nu={1\over4}\vp(1-4L^2_\vp)~. \lb{corol}
\eea

Until now we did not touch any issues related to self-dualities. A link
with the consideration in the previous Section is established by Eq.
\p{imprel} which relates the functions $P_\ab(F)$ and $V_\ab(F)$.

Substituting this into the $SO(2)$ duality condition \p{sdI} and
making use of eq. \p{corol} we find
\bea
&&S[F,P(F)]=[F^\ab - V^\ab(F)]V_\ab(F)
-\mbox{c.c}\nn\\ &&=\nu E_\nu-\bnu E_\bnu=0~. \lb{newsd}
\eea

The corresponding realization of the $SO(2)$ (or $U(1)$) transformations
\p{Ftrans} in terms of $F^\ab$ and $V^\ab(F)$ is given by
\bea
&&\delta_\omega V_\ab=-i \omega V_\ab~,\lb{Vtrans}\\
&&\delta_\omega F_\ab =i \omega [F_\ab -2V_\ab]~.\lb{Ftrans1}
\eea
It should be stressed that the duality transformation realizes linearly
on the auxiliary fields of this representation.

It is important to emphasize that the new form \p{newsd} of the
self-duality constraint \p{sdI} admits a transparent interpretation as
the condition of invariance of $E(\nu,\bnu)$ with respect to the
$U(1)$ transformations \p{Vtrans}
\bea
&& \delta_\omega E = 2i\omega (\bnu E_\bnu - \nu E_\nu) = 0~.
\eea

The general solution of \p{newsd} is a function $\tilde{E}(a)$ which
depends on the single real $U(1)$ invariant variable $a=\nu\bnu$ quartic
in the  auxiliary fields $V_\ab$ and $\bar{V}_{\da\db}$
\bea
&& E_{sd}(\nu, \bnu) = \tilde{E}(a) = \tilde{E}(\nu\bnu)~.
\label{condfin}
\eea
Thus we come to the notable result that the {\it whole} class of nonlinear
extensions of the Maxwell action admitting the on-shell $SO(2)$ duality
is parametrized by an arbitrary $SO(2)$ invariant real function of
one argument $a=V^2\bar V^2$ 
\be
L_{int}^{sd}(F^2,\bar F^2)=\tilde E(a)+a(V^2+\bar V^2)\left(
\frac{d\tilde{E}}{da}\right)^2~.
\ee

Finally, let us examine which restrictions on the interaction Lagrangian
$E(\nu, \bnu)$ are imposed by the requirement of the ``discrete''
self-duality with respect to the Legendre transform.
For this we shall need a first-order representation of the Lagrangian
\p{auxfree}
\bea
&&\hat{L}_2(V,P)\equiv \cL_2(V,F)+iF^\ab P_\ab-i\bar F^{\da\db}
\bar P_{\da\db}={1\over2}(P^2+\bar P^2)
-V^2-\bar V^2\nn\\
&&-2iP^\ab V_\ab+2i\bar P^{\da\db}\bar V_{\da\db}=\cL_2(iV,P)~,\q 
P_\ab=i(F_\ab-2V_\ab)~.
\eea
Thus the Legendre transform   of the quadratic Lagrangian $\cL_2(V,F)$ in 
the variables $F_\ab$ corresponds to the discrete $U(1)$ transformation 
$V_\ab\rightarrow iV_\ab\equiv U_\ab$. The Legendre transform of the whole
Lagrangian \p{legact} has the following form:
\be
\hat{L}(V,P)=\cL_2(U,P)+E(-U^2,-\bar U^2)~.
\ee

Comparing this dual Lagrangian  with the original one
\p{legact}, we observe that the necessary and sufficient condition of the
discrete self-duality is the following simple restriction on the function
$E$ \cite{IZ}
\bea
&& E(\nu,\bnu)=E(-\nu,-\bnu)~. \lb{discr1}
\eea

\section{$U(n)$ self-duality}

Let us consider $n$ Abelian field-strengths
\bea
&& F^i_\ab~,\q \bar{F}^i_{\da\db}~,
\eea
where  $i=1,2\ldots n$. As the first step, one can realize the group
$SO(n)$ on these variables
\bea
&& \delta_\omega F^i_\ab=\xi^{ik}F^k_\ab~,\q \xi^{ki}=-\xi^{ik}~.\lb{On}
\eea
This group is assumed to define an off-shell symmetry of the corresponding nonlinear
Lagrangian $L(F^k, \bar F^k) = -{1\over 2}(F^iF^i)
-{1\over 2}(\bar{F}^i\bar{F}^i) + L_{int}(F^k, \bar{F}^k)$.

The $U(n)$ self-duality conditions for the Lagrangian $L(F^k,\bar{F}^k)$
generalizing the $U(1)$ condition \p{sdI} have been analyzed
in Refs.\cite{ABMZ,KT}. In the spinor notation, these conditions read
\bea
&\cA^{[kl]}=(F^kP^l)-(F^lP^k)-\mbox{c.c.}
=0~,&\lb{Acond}\\
&\cS^{(kl)}=(F^kF^l)+(P^kP^l)-\mbox{c.c.}
=0~,&\lb{Scond}
\eea
where
\bea
&&P^k_\ab(F)\equiv i{\partial L\over\partial F^{k\ab}}~,\q \bar P^k_{\da\db}
\equiv-i{\partial L\over\partial\bar F^{k\da\db}}~,
\\
&&(F^kP^l)\equiv F^{k\ab}P^l_\ab~, \q (\bar F^k\bar P^l)\equiv 
\bar F^{k\da\db}\bar P^l_{\da\db}~ \mbox{etc}~. \nn
\eea

The condition \p{Acond} amounts to the $SO(n)$ invariance of the
Lagrangian and holds off shell. The second condition is the true analog
of \p{sdI}. It guarantees the covariance of the equations of motion for
$F^{k}_\ab,  \bar F^{k}_{\da\db}$ together with Bianchi identities
under the  following nonlinear transformations:
\bea
&&\delta_\eta F_\ab^k=-\eta^{kl}P^l_\ab(F)~,\lb{Btran}\\
&&\delta_\eta P_\ab^k(F)=\eta^{kl}F^l_\ab\nn
\eea
where $\eta^{kl}=\eta^{lk}$ are real parameters. On the surface of the
condition \p{Scond} these transformations, together with \p{On}, form the 
group $U(n)$. The particular solution of the $U(n)$ self-duality conditions 
\p{Acond},\p{Scond} constructed so far \cite{ABMZ}.

The $U(n)$ group structure becomes manifest
after passing to  new auxiliary variables:
\bea
&&V^k_\ab(F)\equiv {1\over2}[F^k_\ab+iP^k_\ab(F)]~,\q\delta V^k_\ab(F)=
\omega^{kl}V^l_\ab(F)~,\lb{Htran}\\
&&\omega^{kl}=\xi^{kl}+i\eta^{kl}~,\q \bar\omega^{lk}=-\omega^{kl}~.\nn
\eea

Passing to an analog of the $V, F$ representation in the $U(n)$ case
will allow us to find the {\it general} solution to \p{Acond}, \p{Scond}.

Let us define a new  representation for the $SO(n)$ invariant nonlinear
electrodynamics Lagrangians in terms of the Abelian gauge field strengths  
$F^k_\ab(A^k)$ and auxiliary fields $V^k_\ab$
\bea
&& \cL(V^k,F^k)=(V^kV^k)+(\bar{V}^k\bar{V}^k)  - 2(V^kF^k)-2(\bar{V}^k
\bar{F}^k)\nn\\
&& +{1\over2}(F^kF^k)  +{1\over2}(\bar{F}^k\bar{F}^k) + E(V^k,
\bar{V}^k)~. \lb{UVF}
\eea
The real Lagrangian of interaction $E(V, \bar V)$ is $SO(n)$ invariant
by definition. 

In the general case of $E\neq 0$ the algebraic equation for $V_\ab$ is
\bea
&& F^k_\ab=V^k_\ab+{1\over2}\frac{\partial E}{\partial V_k^\ab}
~.
\lb{FVrel}
\eea
Using the relation
\bea
 && P^k_\ab=i[F^k_\ab-2V^k_\ab]~,
\eea
one can rewrite the $U(n)$ self-duality conditions \p{Acond} and
\p{Scond} in this representation as follows
\bea
&& i(F^lV^k)-i(F^kV^l)-\mbox{c.c.}=0~, \nn \\
&& (F^lV^k)+(F^kV^l)-2(V^kV^l)-\mbox{c.c.}=0~. \lb{sdn}
\eea
One can readily show that, after making use of the relation \p{FVrel},
these conditions can be brought into the form quite similar to the $U(1)$
self-duality condition \p{newsd}
\bea
&& V^k_\ab\frac{\partial E}{\partial V_\ab^l}-
\bar{V}^k_{\da\db}\frac{\partial E}{\partial\bar{V}_{\da\db}^l}=0~.
\lb{EUinvar}
\eea
This constraint is none other than the condition of invariance of
$E(V, \bar V)$ with respect to the $U(n)$ transformations \p{Htran}.
The simplest example of the $U(n)$ self-dual action is defined by
the  quartic interaction of the auxiliary fields
\be
E=(V^kV^l)(\bar V^k\bar V^l)~,\q F^k_\ab=V^k_\ab+V^l_\ab(\bar V^k\bar V^l)~.
\ee
The corresponding non-polynomial Lagrangian $L(F^k,\bar F^k)$ can be 
constructed from \p{UVF} using  the  solution $V^k(F,\bar F)$ of the 
last algebraic equation.

Finally, let us notice that the condition of ``discrete'' self-duality in
the general case is as follows
\bea
&& E(V^k_\ab, \bar V^k_{\da\db}) = E(iV^k_\ab, -i \bar V^k_{\da\db})~.
\lb{discrn}
\eea
It is an obvious generalization of the $n=1$ condition \p{discr1}.

This work was partially supported by 
INTAS grant No 00-254, RFBR-DFG grant No 02-02-04002, grant DFG 436 RUS 
113/669 and a grant of Heisenberg-Landau
Programme.


\begin{thebibliography}{9}
\bibitem{IZ2} E.A. Ivanov and B.M. Zupnik, \textit{New representation for 
Lagrangians of self-dual nonlinear electrodynamics}, Proceedings of XVI Max 
Born Symposium "Supersymmetries and quantum symmetries", eds. E. Ivanov et 
al,  p. 235, Dubna (2002); hep-th/0202203.
\bibitem{IZ} E.A. Ivanov and B.M. Zupnik, \textit{N=3 supersymmetric 
Born-Infeld theory}, Nucl. Phys. B 618 (2001) 3; hep-th/0110074.
\bibitem{GZ}M.K. Gaillard and B. Zumino,
\textit{Nonlinear electromagnetic self-duality and
Legendre transformation},  Duality and Supersymmetric
Theories, eds. D.I. Olive et al, p. 33, Cambridge University
Press, 1999; hep-th/9712103.
\bibitem{GR} G.W. Gibbons and D.A. Rasheed, \textit{ Electric-magnetic 
duality rotations in non-linear electrodynamics}, Nucl. Phys. B 454
(1995) 185; hep-th/9506035.
\bibitem{KT} S.M. Kuzenko and S. Theisen,
\textit{Nonlinear self-duality and supersymmetry}, Fortsch. Phys. 49 (2001) 
 273;  hep-th/0007231. 
\bibitem{HKS}M. Hatsuda, K. Kamimura and S. Sekia,
\textit{ Electric-magnetic duality invariant Lagrangians}, 
Nucl.Phys. B 561  (1999) 341 ; hep-th/9906103.
\bibitem{ABMZ} P. Aschieri, D. Brace, B. Morariu and B. Zumino, \textit{
Nonlinear selfduality in even dimensions}, Nucl. Phys.
B 574 (2000) 551;  hep-th/9909021.
\end{thebibliography}
\end{document}